# Crystal, magnetic, and electronic structures, and properties of new BaMn$Pn$F ($Pn$ = As, Sb, Bi)

Bayrammurad Saparov*,[1], David J. Singh[1], V. Ovidiu Garlea[2], Athena S. Sefat[1]

[1] Materials Science and Technology Division, Oak Ridge National Laboratory, Oak Ridge, TN 37831, USA
[2] Quantum Condensed Matter Division, Oak Ridge National Laboratory, Oak Ridge, TN 37831, USA

**ABSTRACT:** New fluoropnictides BaMn$Pn$F with $Pn$ = As, Sb, Bi, are synthesized by stoichiometric reaction of elements with BaF$_2$. The compounds crystallize in the tetragonal $P4/nmm$ (No. 129, Z = 2) space group, with the ZrCuSiAs-type structure, as indicated by single crystal and powder X-ray diffraction results. Electrical resistivity results indicate that $Pn$ = As, Sb, and Bi are semiconductors with band gaps of $E_g$ = 0.73 eV, $E_g$ = 0.48 eV and $E_g$ = 0.003 eV, respectively. Powder neutron diffraction reveals a $G$-type antiferromagnetic order below $T_N$ = 338(1) K for $Pn$ = As, and below $T_N$ = 272(1) K for $Pn$ = Sb. Magnetic susceptibility increases with temperature above 100 K for all the materials. Density functional calculations also find semiconducting antiferromagnetic compounds with strong in-plane and weaker out-of-plane exchange coupling that may result in non-Curie Weiss behavior above $T_N$. There is strong covalency between Mn and pnictogen elements. The ordered magnetic moments are 3.65(5) $\mu_B$/Mn for $Pn$ = As, and 3.66(3) $\mu_B$/Mn for $Pn$ = Sb at 4 K, as refined from neutron diffraction experiments.

## INTRODUCTION

The recent discovery of high temperature superconductivity in F-doped LaFeAsO[1] (1111 family) initiated an extensive research into analogous materials. This research lead to the discoveries of superconductivity in doped BaFe$_2$As$_2$,[2,3] (122 family), LiFeAs[4] (111 family), and FeSe[5] (11 family), among others. All these families feature two dimensional (2D) structures with FeAs or FeSe layers, which contain edge-shared FeAs$_4$ or FeSe$_4$ tetrahedra. Fe atoms are formally divalent; hole-, electron-, or isovalent-doping inside or outside of layers can result in superconductivity.

Among the Fe-based superconductors (FeSCs), the highest superconducting transition temperatures ($T_C$s) are reported for the 1111 family.[6] In the search for non-Fe-based oxypnictides, varieties of physical properties are found such as diamagnetism in LaZnAsO,[7] itinerant ferromagnetism in LaCoAsO,[8,9] semiconducting antiferromagnetism in $Ln$Mn$Pn$O[10] ($Ln$ = La-Sm; $Pn$ = P, As) and superconductivity in LaNiAsO[11] ($T_C$ ~ 3 K). Interestingly, although such oxyarsenides have been studied in detail, reports on non-oxide 1111 fluoroarsenides with ZrCuSiAs-type structure, i.e. $B$FeAsF ($B$= alkaline-earth metal), are relatively scarce.

For the 1111 fluopnictides $B$FeAsF, rare-earth substitution at the alkaline-earth metal ($B$) site is found to give the highest $T_C$ of 56 K for Sr$_{0.5}$Sm$_{0.5}$FeAsF and Ca$_{0.4}$Nd$_{0.6}$FeAsF.[12,13] The only known non-Fe transition-metal-based 1111 fluoropnictides are BaMnPF,[14] $A$Cu$Ch$F and $A$Ag$Ch$F ($A$ = Sr, Ba, Eu; $Ch$ = S, Se, Te).[15,16] However, it should be noted that Cu and Ag in $A$Cu$Ch$F and $A$Ag$Ch$F are nominally monovalent, and therefore, BaMnPF is the only known 1111 fluoropnictide containing divalent non-Fe transition metal. Other known 1111 fluoropnictides, include those of group 12 metals, namely SrZnPF, BaZnPF, BaZnSbF and BaCdSbF.[14,17] According to literature on the properties of 1111 fluoropnictides, SrCu$Ch$F,[18,19] BaCu$Ch$F[20] and EuCu$Ch$F[19] ($Ch$ = S, Se and Te) are $p$-type semiconductors with Seebeck coefficients ranging from +10 to +620 μV/K. BaZnPF[14] and BaCdSbF[17] are also semiconductors with band gaps of $E_g$ = 0.5 eV and 0.25 eV, respectively. Moreover, BaMnPF is a semiconducting antiferromagnet with a temperature independent magnetic susceptibility up to 300 K.[14]

Considering the fact that doped $B$FeAsF fluoropnictides are among the highest $T_C$ (56 K) FeSCs, there is an incentive to explore for superconductivity in similar 1111 structures, and even non-Fe-based analogs. This is the report of the synthesis of new 1111 ZrCuSiAs-type BaMn$Pn$F ($Pn$ = As, Sb, Bi) (Figure 1) and a comprehensive study of their crystal, magnetic, and electronic structures. We report nuclear structures from powder and single crystal X-ray diffraction, and magnetic structures from neutron powder diffraction. In addition, we present thermodynamic and transport properties from temperature dependent electrical transport data, temperature- and field-dependent magnetization data. Moreover, we report density functional theory (DFT) electronic structure calculations.

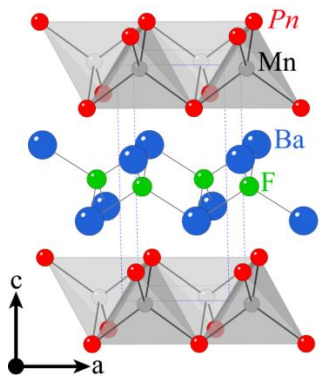

**Figure 1.** Tetragonal ZrCuSiAs-type structure of BaMn*Pn*F (*Pn* = As, Sb, Bi). The unit cell contains transition metal pnictide layers of [Mn*Pn*]⁻ that are made of Mn*Pn*$_4$ tetrahedra.

## EXPERIMENTAL SECTION

**Synthesis.** Dendritic Ba, Mn and As pieces, and Sb and Bi granules with purities greater than 99.9% are used as received from Alfa Aesar. Ultrapure BaF$_2$ powder, from Ventron Alfa Inorganics, is dried at 200°C for 3 h before using. Reactants in the stoichiometric ratio of Ba:BaF$_2$:Mn:*Pn* = 1:1:2:2 are weighed inside a helium-filled glovebox and put into alumina crucibles. The alumina crucibles are then transferred into silica tubes and sealed under vacuum. The reaction mixtures are heated to 1000°C (30°C h⁻¹; dwell 6 h), then to 900°C (10°C h⁻¹; dwell 6 h), and subsequently to 300°C (30°C h⁻¹) after which the furnaces are switched off. This initial sintering step produced multi-phase reaction products, along with μm-size crystallites of 1111 phase in *Pn* = Sb and Bi, which are extracted for structural determinations using single crystal X-ray diffraction. The products are reground and pelletized inside the glovebox. The pellets are then placed inside alumina crucibles, enclosed in silica tubes, vacuum sealed, then annealed for a second time at 900°C (dwell 60 h). The third sintering step is the repeat of the latter annealing procedure. This only results in marginal improvement of phase purity as judged by slightly lower BaF$_2$ impurity levels compared to the second sintering step.

**Characterization**

*X-ray Diffraction.* For BaMn*Pn*F (*Pn* = As, Sb, Bi), powder X-ray diffraction (PXRD) data are collected on a PANalytical X'Pert PRO MPD X-ray Diffractometer using monochromated Cu-K$\alpha_1$ radiation. Scans are performed in 5-65° (2θ) range, with a step size of 1/60° and 20-100 seconds/step counting time. Low temperature data collections are carried out using an Oxford Phenix closed cycle cryostat. Due to the air sensitivity of finely ground powders of *Pn* = Sb and Bi, the powders are loaded in a protective holder with a polycarbonate cover, inside the glovebox. Rietveld refinements are performed using GSAS[21,22] software package. PXRD results are summarized in Figures 2 to 4, and Table S1 (see Supporting Information).

For *Pn* = Sb and Bi, single crystal x-ray diffraction data are collected on a Bruker SMART APEX CCD-based diffractometer, which employs Mo K$\alpha$ radiation ($\lambda$ = 0.71073 Å). Crystals are selected under a microscope and cooled to 173(2) K under a cold nitrogen stream. The data collection, data integration and refinement are carried out using the SHELXTL software package.[23] SADABS is used for semi-empirical absorption correction based on equivalents. The structures are solved by direct methods and refined by full matrix least-squares methods on F². All sites are refined with anisotropic atomic displacement parameters and full occupancies. Details of the crystallographic data and structure refinement parameters are given in Table 1. Positional and equivalent isotropic displacement parameters, along with refined interatomic distances and angles are provided in Tables 2 and 3, respectively. Further details of the crystal structure studies are summarized in the form of CIF (Crystallographic Information File) file (see Supporting Information).

*Elemental analysis.* Energy-dispersive X-ray spectroscopy (EDS) measurements are carried out using a Hitachi-TM3000 scanning electron microscope equipped with a Bruker Quantax 70 EDS system. Data acquisition is carried out with an accelerating voltage of 15 kV in 2-3 min scanning time. EDS results confirm elemental compositions of 1:1:1:1 for BaMnAsF, BaMnSbF and BaMnBiF.

*Physical property measurements.* For BaMn*Pn*F (*Pn* = As, Sb, Bi), DC magnetization measurements are carried out using a Quantum Design Magnetic Property Measurement System. Temperature dependence of magnetization experiments are performed on polycrystalline pellets under applied fields of 10 kOe, 20 kOe and 30 kOe. Both zero field cooled (ZFC) and field cooled (FC) data are collected. Magnetization measurements as a function of field are carried out at 5 K, 100 K and 300 K. Four-probe electrical resistivity measurements are carried out on a Quantum Design Physical Property Measurement System.

*Thermal analysis.* Thermogravimetric analysis (TGA) and differential thermal analysis (DTA) on BaMn*Pn*F (*Pn* = As, Sb) are performed using a Pyris Diamond TG/DTA from Perkin Elmer, under a stream of ultra-high purity argon gas. The measurements are carried out on 20-25 mg pellet pieces in the temperature interval of 323-1573 K (20 K min⁻¹).

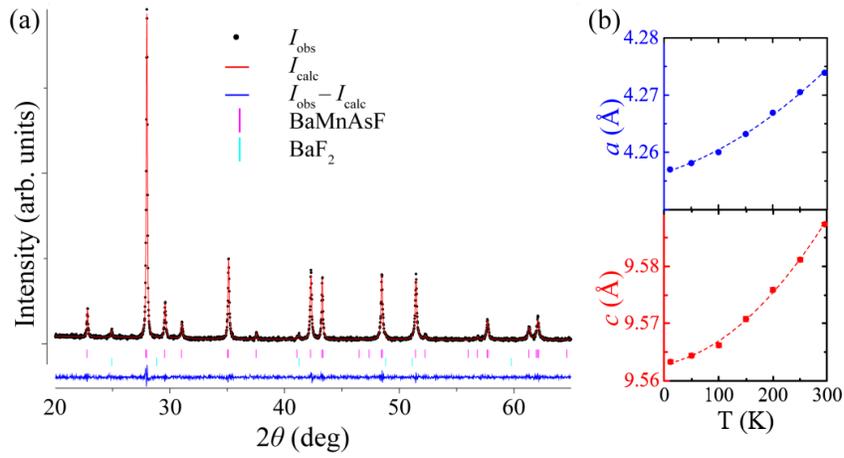

**Figure 2**. For BaMnAsF, (a) Rietveld refinement (red line) of powder X-ray diffraction data (black dots), shown along with Bragg positions for 1111 tetragonal $I4/mmm$ phase (pink tick marks) and $BaF_2$ impurity (turquoise tick marks); (b) temperature-dependence of refined lattice parameters (dash lines are guide to the eye).

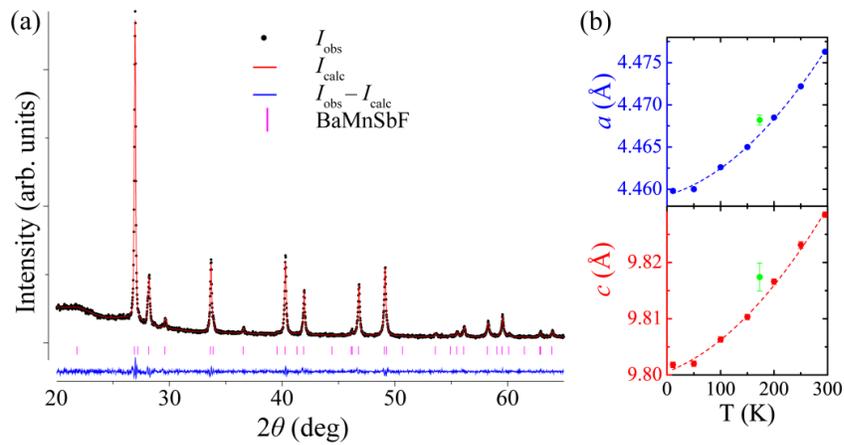

**Figure 3**. For BaMnSbF, (a) Rietveld refinement (red line) of powder X-ray diffraction data (black dots), shown along with Bragg positions for 1111 tetragonal $I4/mmm$ phase (pink tick marks) with no impurities; (b) temperature-dependence of refined lattice parameters (dash lines are guide to the eye). The green data points in (b) are $a$- and $c$- lattice parameters from single crystal X-ray diffraction refinements.

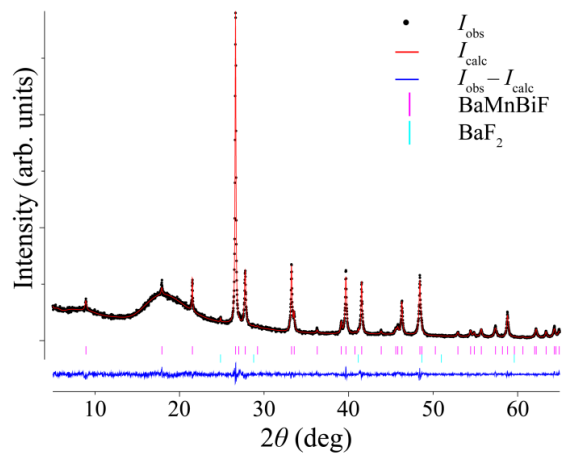

**Figure 4**. For BaMnBiF, (a) Rietveld refinement (red line) of powder X-ray diffraction data (black dots), shown along with Bragg positions for 1111 tetragonal $I4/mmm$ phase (pink tick marks) and $BaF_2$ impurity (turquoise tick marks).

*Neutron Powder Diffraction.* Neutron powder diffraction experiments on BaMn*Pn*F (*Pn* = As and Sb) are carried out using the HB-2A neutron powder diffractometer at the High Flux Isotope Reactor (HFIR) at Oak Ridge National Laboratory (ORNL). Measurements are performed using two different wavelengths of λ = 1.536 Å and 2.410 Å provided by the (115) and (113) reflections of a vertically focusing Ge monochromator; this allows for optimization of the instrument resolution function for specific Q ranges. The data are collected by scanning the detector array consisting of 44 $^3$He tubes in two segments to cover the total 2θ range of 6-150° in steps of 0.05°; overlapping detectors for the given step average the counting efficiency of each detector.[24] For the measurements, 4 g powder samples are confined inside vanadium containers. For below room temperature measurements, samples are loaded inside a JANIS top-loading closed-cycle refrigerator, while for high temperatures experiments up to 800 °C, samples are loaded in an ILL vacuum furnace equipped with Nb heating elements. Rietveld refinements are performed using the FULLPROF program.[25] Spin configurations compatible with the crystal symmetry are generated by group-theory analysis using the program SARAh.[26] In *Pn* = As compound, impurity phases of BaF$_2$ (1.5% weight fraction) and Mn$_2$As (2.5%) are noted; in *Pn* = Sb compound, BaF$_2$ (2.5% weight fraction) and Mn$_{1-x}$O (2.4%) impurities are found.

*Electronic structure calculations.* First principles calculations are performed using the experimental crystal structure information data. The calculations are done within DFT using the generalized gradient approximation (GGA) of Perdew, Burke and Ernzerhof[27] and the general potential linearized augmented planewave (LAPW) method[28] as implemented in the WIEN2k code.[29] Well-converged basis sets and Brillouin zone samplings are employed, along with LAPW sphere radii of 2.4 Bohr for Ba, Mn, As, Sb and Bi, and 1.9 Bohr for F. Local orbitals[30] are added to the basis to include semi-core states and spin-orbit is included in the calculations.

## RESULTS AND DISCUSSION

**Synthesis, crystal chemistry, and stability**. Three new compounds of BaMnAsF, BaMnSbF and BaMnBiF are synthesized using the solid-state sintering method. The compounds crystallize in the primitive ZrCuSiAs-type tetragonal *P*4/*nmm* (No. 139, Pearson symbol *tP*8), and are isotypic to the lighter BaMnPF (Figure 1).[14]

EDS results confirm the presence of small crystallites with BaMnSbF and BaMnBiF compositions after the first heating step, which allows for collection of single-crystal x-ray diffraction data (Table 1). There are four crystallographically unique atoms in the asymmetric unit cell, all located in special positions (Tables 2, 4, S1). BaMn*Pn*F structure, similar to LaFeAsO, can be viewed with cationic $^2_\infty$[BaF]$^+$ and anionic $^2_\infty$[Mn*Pn*]$^-$ layers alternating along the c-axis. These layers are built upon edge-sharing FBa$_4$ and Mn*Pn*$_4$ tetrahedra, respectively. Following the discovery of superconductivity in the F-doped LaFeAsO, the ZrCuSiAs-type structure and its relationship with other structures have been extensively studied.[10,31,32]

From single crystal X-ray data (Table 3), Mn-Sb and Mn-Bi bond distances in the [Mn*Pn*]$^-$ layers are $d_{\text{Mn-Sb}}$ = 2.7767(5) Å and $d_{\text{Mn-Bi}}$ = 2.8479(11) Å, respectively; the Mn-Mn distances are $d_{\text{Mn-Mn}}$ = 3.1595(4) Å in BaMnSbF and $d_{\text{Mn-Mn}}$ = 3.196(1) Å in BaMnBiF. The tetrahedral angles are 110.65(1)° and 107.14(3)° in MnSb$_4$, and 111.73(3)° and 105.04(5)° in MnBi$_4$ tetrahedra. The interlayer distances shown by the distance between barium and pnictogen atoms in adjacent layers are $d_{\text{Ba-Sb}}$ = 3.6489(6) Å in BaMnSbF and $d_{\text{Ba-Bi}}$ = 3.671(1) Å in BaMnBiF. For comparison, MnAs$_4$ tetrahedra in BaMnAsF are characterized by the distances $d_{\text{Mn-As}}$ = 2.60460(4) Å and $d_{\text{Mn-Mn}}$ = 3.0221(1) Å, and bond angles 109.0789(8)° and 110.259(2)°, from the room temperature PXRD data (see Table S1). The bond distances and angles in BaMn*Pn*F are very close to those reported for the related BaMn$_2$*Pn*$_2$ ternary phases,[33-35] which also contain anionic $^2_\infty$[Mn*Pn*]$^-$ layers.

**Table 1** For BaMn*Pn*F (*Pn* = Sb, Bi), selected single crystal X-ray diffraction data and structure refinement parameters.

| Empirical formula | BaMnSbF | BaMnBiF |
|---|---|---|
| Temperature, K | 173(2) | |
| Radiation, Å | Mo K$\alpha$, 0.71073 | |
| Crystal system | Tetragonal | |
| Space group, Z | *P*4/*nmm* (No. 129), 2 | |
| Formula weight, g mol$^{-1}$ | 333.03 | 420.26 |
| *a*, Å | 4.4682(6) | 4.5198(14) |
| *c*, Å | 9.817(3) | 9.875(7) |
| *V*, Å$^3$ | 196.00(6) | 201.74(16) |
| $\rho_{\text{calc}}$, g cm$^{-3}$ | 5.643 | 6.918 |
| $\mu$, mm$^{-1}$ | 19.736 | 56.034 |
| $\theta_{\min} - \theta_{\max}$, ° | 2.07-28.20 | 2.06-28.50 |
| $R_1^a$ (all data) | 0.0248 | 0.0388 |
| $wR_2^a$ (all data) | 0.0644 | 0.0963 |
| Goodness-of-fit on $F^2$ | 1.23 | 1.11 |
| Largest peak and hole, e$^-$ Å$^{-3}$ | 3.35 and -0.95 | 3.18 and -1.79 |

$^aR_1 = \sum \left\lVert F_o \right\rvert - \left\lvert F_c \right\rVert / \sum \left\lvert F_o \right\rvert; wR_2 = \left\lVert \sum \left\lvert w(F_o^2 - F_c^2)^2 \right\rvert / \sum \left\lvert w(F_o^2)^2 \right\rvert \right\rVert^{1/2}$ and $w = 1 / \left\lvert \sigma^2 F_o^2 + (AP)^2 + BP \right\rvert$, and $P = (F_o^2 + 2F_c^2)/3$; *A* and *B* are weight coefficients.

**Table 2** For BaMn*Pn*F (*Pn* = Sb, Bi), atomic coordinates and equivalent isotropic ($U_{eq}{}^a$) displacement parameters, at 173 K, from single crystal X-ray diffraction refinements.

| Atom | Wyckoff site | x | y | z | $U_{eq}$ (Å$^2$) |
|---|---|---|---|---|---|
| | | BaMnSbF | | | |
| Ba | 2c | 0.25 | 0.25 | 0.3539(1) | 0.0113(3) |
| Mn | 2a | 0.25 | 0.75 | 0 | 0.0117(4) |
| Sb | 2c | 0.25 | 0.25 | 0.8321(1) | 0.0108(3) |
| F | 2b | 0.25 | 0.75 | 0.5 | 0.013(1) |
| | | BaMnBiF | | | |
| Ba | 2c | 0.25 | 0.25 | 0.3585(2) | 0.0205(5) |
| Mn | 2a | 0.25 | 0.75 | 0 | 0.0210(8) |
| Bi | 2c | 0.25 | 0.25 | 0.8245(1) | 0.0205(4) |
| F | 2b | 0.25 | 0.75 | 0.5 | 0.023(3) |

$^a U_{eq}$ is defined as one third of the trace of the orthogonalized $U_{ij}$ tensor.

**Table 3** For BaMn*Pn*F (*Pn* = Sb, Bi), selected bond distances (Å) and angles (°), at 173 K, from single crystal X-ray diffraction.

| BaMnSbF | | BaMnBiF | |
|---|---|---|---|
| F–Ba (×4) | 2.6549(5) | F–Ba (×4) | 2.6572(12) |
| Mn–Sb (×4) | 2.7767(5) | Mn–Bi (×4) | 2.8479(11) |
| Mn–Mn | 3.1595(4) | Mn–Mn | 3.1960(10) |
| Ba–Sb | 3.6489(6) | Ba–Bi | 3.6714(14) |
| Sb–Mn–Sb | 110.65(1) | Bi–Mn–Bi | 111.73(3) |
| Sb–Mn–Sb | 107.14(3) | Bi–Mn–Bi | 105.04(5) |
| Ba–F–Ba | 114.60(3) | Ba–F–Ba | 116.53(7) |
| Ba–F–Ba | 106.97(1) | Ba–F–Ba | 106.06(3) |

PXRD patterns along with Rietveld refinements results for BaMnAsF, BaMnSbF and BaMnBiF are illustrated in Figures 2 to 4. There are no Bragg peaks in the 2θ range of 5-20° for *Pn* = As and Sb, and so the low angle regions are not shown for them. The broad background below 2θ ~ 25° in *Pn* = Sb and Bi (Figure 3-4) are caused by the polycarbonate cover. X-ray diffraction refinement detects a small amount of BaF$_2$ crystalline impurity (less than 1% by mass) in *Pn* = As and Bi products, whereas no impurity peaks are observed in the *Pn* = Sb compound.

All BaMn*Pn*F samples have MnO impurities if ground in air during sample sintering stages. Even in the final product form, ground-in-air BaMnSbF shows signs of oxidation in the PXRD data, while BaMnBiF burns in air. The extreme air sensitivity of ground BaMnBiF limits its full characterization through neutron diffraction experiments. Notwithstanding these facts, the heating of pellets under ambient conditions up to 130°C and subsequent PXRD measurements demonstrate that BaMn*Pn*F are air-stable in pellet form.

The refined room temperature lattice parameters from the PXRD data are *a* = 4.2739(1) Å and *c* = 9.5875(2) Å for BaMnAsF (*Rp* = 8.54%, *wRp* = 11.52%), *a* = 4.4791(1) Å and *c* = 9.8297(2) Å for BaMnSbF (*Rp* = 4.59%, *wRp* = 6.03%), and *a* = 4.5384(1) Å and *c* = 9.8929(2) Å for BaMnBiF (*Rp* = 4.05%, *wRp* = 5.40%). The refinement of PXRD data on BaMnAsF and BaMnSbF show (Figures 2b, 3b) that there are no structural anomalies upon cooling and the unit cell volumes contract uniformly. A survey of the Inorganic Crystal Structure Database (ICSD)[15] and the Pearson's Handbook[16] reveals new features about the BaMn*Pn*F family. First, they are the only reported compounds in each of the corresponding Ba-Mn-*Pn*-F (*Pn* = As, Sb, Bi) phase diagrams so far. Second, BaMnBiF can also be considered to be the first bismuthide with the ZrCuSiAs-type structure, considering that Bi in BiCuSeO[36] serves as a cation. Third, together with BaMnPF, the new compounds are the first phases with Mn and F to adopt the ZrCuSiAs structure type. Fourth, the cell volumes of BaMnSbF and BaMnBiF (see Table 1) are the largest within this structure type along with that of BaCdSbF.[17] As expected from the large unit cell volumes and lattice parameters of BaMnSbF and BaMnBiF, the bond distances are also longer than those reported for other compounds that adopt this structure type.[7,10]

The thermal behaviors of BaMnAsF and BaMnSbF are studied through TGA/DTA (Figure S1). Both BaMnAsF and BaMnSbF are stable up to 1300 K (~ 1030 °C), after which they decompose. Additionally, pellets of BaMnAsF and BaMnSbF are separately vacuum sealed inside silica tubes and heated to 1500 K with periodic visual monitoring. This is done because the DTA on BaMnAsF (Figure S1) shows two peaks upon heating, and it is not immediately clear if the first peak corresponds to the melting of the compound. No visual change occurs to the pellets up to ~ 1430 K, after which molten liquid is clearly visible in both samples. These molten pieces contain MnAs and MnSb binaries according to the EDS results.

**Physical properties**. For BaMn*Pn*F, temperature dependence of electrical resistivity results are plotted in Figure 5. The compounds show semiconducting behavior, with room temperature resistivity values of $\rho_{300K}$(BaMnAsF)= 3.6 ×10$^5$ Ω cm, $\rho_{300K}$(BaMnSbF)= 2.4 ×10$^4$ Ω cm, and $\rho_{300K}$(BaMnBiF)= 0.135 Ω cm. Similarly, BaMnPF is also reported as a semiconductor.[14] The semiconducting behavior supports the charge-balanced nature of the compounds according to [Ba$^{2+}$F$^-$][Mn$^{2+}$*Pn*$^{3-}$]. $d\rho/dT$ derivatives of the resistivity data are featureless in the measured range. Calculated band gaps from the Arrhenius fit ($\ln\rho = \ln\rho_0 + E_g/2k_BT$) are $E_g$(BaMnAsF)= 0.73 eV, $E_g$(BaMnSbF)= 0.48 eV, and $E_g$(BaMnBiF)= 0.003 eV (see Figure 5). These band gap values follow the expected periodic trend based on electronegativities of pnictogen elements.[37] It is interesting that for BaMnBiF, both the temperature dependence of electrical resistivity and the calculated narrow band gap are quite different when compared to the other *Pn* members. For BaMnBiF, $\rho(T)$ decreases upon cooling down to ~ 85 K, then changes slope and sharply increases below 80 K. From the trend in the band gaps in this family, it can be speculated that BaMnBiF is an extrinsic semiconductor.[38]

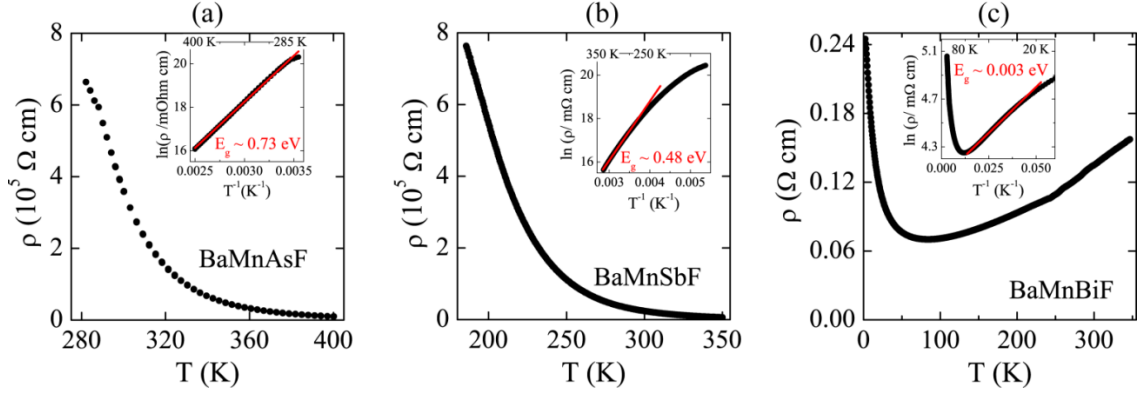

**Figure 5**. Temperature dependence of electrical resistivity, $\rho(T)$, for polycrystalline pellets of (a) BaMnAsF, (b) BaMnSbF and (c) BaMnBiF. Insets illustrate lines (in red) corresponding to Arrhenius fit for the temperature region.

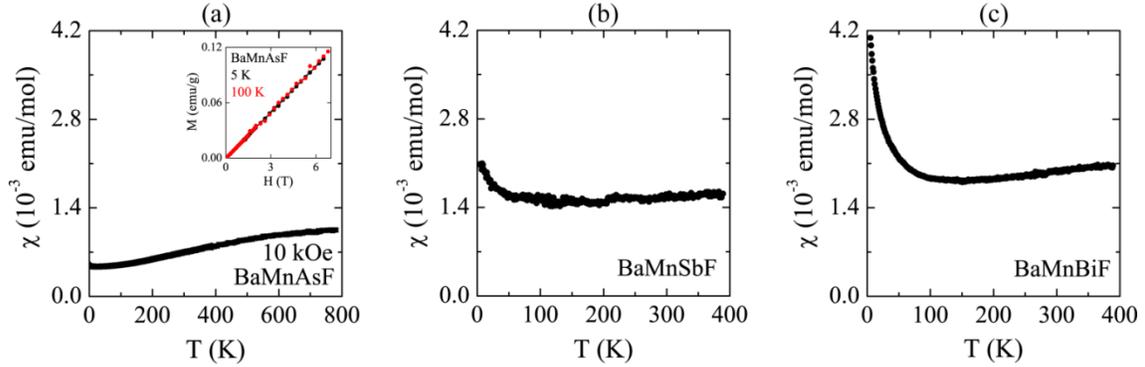

**Figure 6**. Temperature dependence of magnetic susceptibility, $\chi(T)$, for (a) BaMnAsF, (b) BaMnSbF and (c) BaMnBiF. For BaMnAsF, inset shows a linear field-dependence of magnetization, $M(H)$, at 5 K and 100 K. Note that (b) and (c) are derived using magnetization data under 20 kOe and 30 kOe assuming $\chi \approx \Delta M/\Delta H$, in order to eliminate ferromagnetic impurities (see text).

The calculated band gaps for BaMnAsF and BaMnSbF are two orders of magnitude larger than those reported for the narrow gap semiconductors of BaMn$_2$Pn$_2$, with band gaps of 6-54 meV.[33-35] Such a difference may be attributed to the fact that BaMnPnF contain an additional insulating [BaF]$^+$ layer in the structure, which reduces the band dispersion in the c-axis direction. Further comparisons can be made with the isostructural quaternary phases based on manganese. Most of LnMnPnO (Ln = rare-earth metal) compositions have been reported to be semiconductors with varying band gaps,[10] with exceptions such as metallic PrMnSbO.[39]

Temperature- and field-dependent magnetization results for BaMnAsF are plotted in Figure 6a. ZFC and FC $\chi(T)$ data overlap for BaMnAsF, and are measured up to ~800 K. There is an upward tail in $\chi(T)$ below ~ 15 K, which is probably due to paramagnetic impurities, above which magnetic susceptibility increases slowly with rising temperature. There is another change in the slope of $\chi(T)$ at ~ 550 K, above which $\chi(T)$ starts to plateau. $M(H)$ plots both at 5 K and 100 K are linear. $\chi(T)$ and $M(H)$ data, coupled with resistivity data, suggest that BaMnAsF is likely a local moment antiferromagnet.

Field dependence of magnetization (not shown) for BaMnSbF and BaMnBiF suggest presence of ferromagnetic impurities. From a linear fit to the $M(H)$ data at 5 K, saturation magnetization values of $M_{sat}$(BaMnSbF)= 0.039 $\mu_B$/mol Mn, and $M_{sat}$(BaMnBiF)= 0.006 $\mu_B$/mol Mn are obtained. Assuming that the sources of the ferromagnetic impurities are MnSb or MnBi binaries,[15,16] the $M_{sat}$ values correspond to 1.2% (molar) MnSb, or 0.13% MnBi impurity concentrations. Consequently, in order to find intrinsic behavior for BaMnSbF and BaMnBiF compounds, magnetic susceptibility measurements were performed under two applied fields (20 kOe and 30 kOe), then subtracted from each other according to $\chi \approx \Delta M/\Delta H$; the resulting $\chi(T)$ plots are shown in Figure 6b, c. Although there are low temperature upturns below ~ 100 K, $\chi(T)$, similar to BaMnAsF, increase slowly with rising temperature.

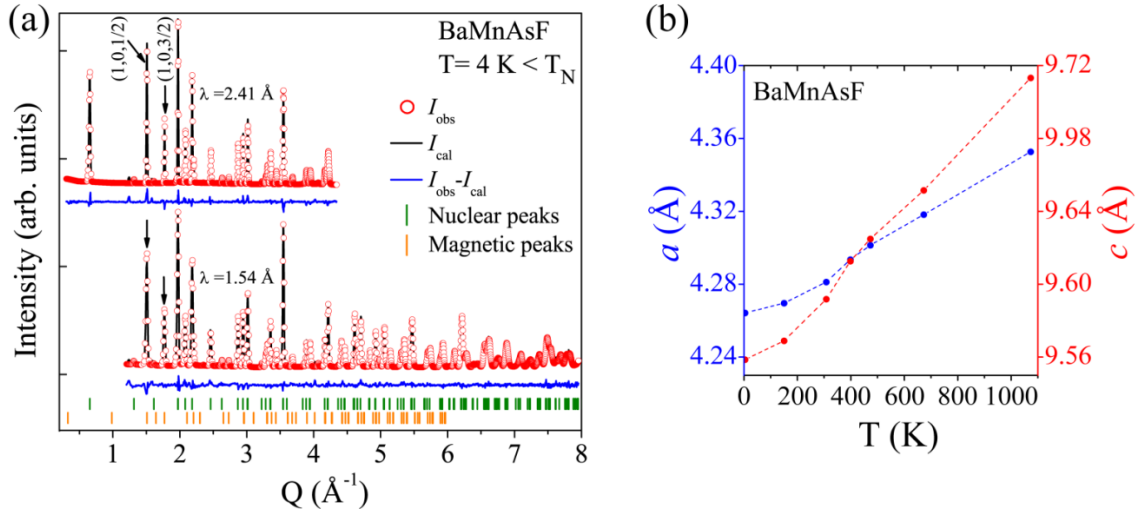

**Figure 7.** (a) Rietveld plots of the neutron powder diffraction data for BaMnAsF collected at 4 K using 1.54 Å and 2.41 Å wavelengths. The two most prominent magnetic reflections, indexed as (1 0 1/2) and (1 0 3/2), are indicated by arrows. (b) Trends in lattice parameters for BaMnAsF as obtained from the NPD data. Dashed lines are intended as a guide to the eye.

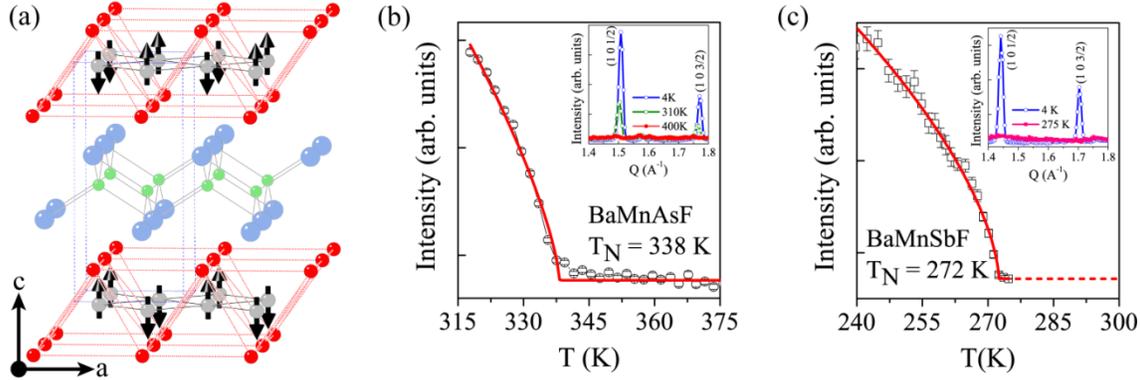

**Figure 8.** (a) The magnetic structure for BaMnAsF and BaMnSbF. Temperature dependence of (1 0 0.5) magnetic peak for (b) BaMnAsF and (c) BaMnSbF. Insets of (b) and (c) show evolution of (1 0 1/2) and (1 0 3/2) magnetic peaks with lowering temperature in low Q region.

The long range antiferromagnetic order found from neutron diffraction results (see next section) at $T_N$ = 338(1) K for BaMnAsF, and $T_N$ = 272(1) K for BaMnSbF do not manifest clearly in the magnetic susceptibility data (Figure 6). Magnetic susceptibility data without a feature at $T_N$ is not unique to BaMnAsF and BaMnSbF; for example, $\chi(T)$ for LaMnPO[40] ($T_N$ = 375(5) K) is featureless up to 800 K; this was attributed to a very strong exchange coupling in the compound. Cases of increasing $\chi(T)$ above $T_N$ have also been reported for $BaMn_2As_2$,[41] $BaMn_2Bi_2$,[35] $BaFe_2As_2$[3] and LaFeAsO.[42]

**Neutron powder diffraction.** Results of the neutron powder diffraction (NPD) experiments are summarized in Table 4, and Figures 7 and 8, and S2 to S4. Rietveld refinement of the NPD data at 4 K gives lattice parameters of $a$ = 4.2642(1) Å and $c$ = 9.5586(3) Å for BaMnAsF ($R_{Bragg}$ = 3.54%, $R_F$ = 2.4%), and $a$ = 4.4636(1) Å and $c$ = 9.7885(4) Å for BaMnSbF ($R_{Bragg}$ = 4.85%, $R_F$ = 3.2%). Magnetic peaks can be indexed by a wave vector **k** = (0 0 1/2), giving a magnetic structure (Figure 8a) with antiferromagnetic (AF) coupling between nearest-neighbor Mn ions in the [Mn$Pn$]$^-$ layers. The adjacent [Mn$Pn$]$^-$ layers are also antiferromagnetically coupled, leading to a $G$-type AF order and a magnetic unit cell twice the size of the chemical unit cell. The NPD data at 4 K yield ordered moment values of 3.65(5) $\mu_B$/Mn for BaMnAsF and 3.66(3) $\mu_B$/Mn for BaMnSbF, which are lower than the value of 5.00 $\mu_B$/Mn expected for the $S$ = 5/2 (high spin) Mn$^{2+}$ ions, assuming $\mu$ = $gS\mu_B$ with $g$ = 2.

**Table 4** Refined structural parameters for BaMnAsF and BaMnSbF at 4 K from neutron powder diffraction, in $P4/nmm$.

| Atom | Wyckoff site | $x$ | $y$ | $z$ | $B$ (Å$^2$) |
|---|---|---|---|---|---|
| | | BaMnAsF | | | |
| | | $a = 4.2642(1)$ Å, $c = 9.5586(3)$ Å | | | |
| Ba | 2$c$ | 0.25 | 0.25 | 0.3386(5) | 0.09(8) |
| Mn | 2$a$ | 0.25 | 0.75 | 0 | 0.12(7) |
| As | 2$c$ | 0.25 | 0.25 | 0.8459(3) | 0.12(6) |
| F | 2$b$ | 0.25 | 0.75 | 0.5 | 0.29(6) |
| | | BaMnSbF | | | |
| | | $a = 4.4636(1)$ Å, $c = 9.7885(4)$ Å | | | |
| Ba | 2$c$ | 0.25 | 0.25 | 0.3546(4) | 0.05(7) |
| Mn | 2$a$ | 0.25 | 0.75 | 0 | 0.29(8) |
| Sb | 2$c$ | 0.25 | 0.25 | 0.8316(4) | 0.15(7) |
| F | 2$b$ | 0.25 | 0.75 | 0.5 | 0.48(6) |

Although lower than expected, the ordered moment values of 3.65(5) $\mu_B$/Mn for BaMnAsF, and 3.66(3) $\mu_B$/Mn for BaMnSbF suggest local moment antiferromagnetism in these compounds. In comparison, the ordered moment values are 0.2-1.0 $\mu_B$/Fe in the delocalized spin-density-wave (SDW) antiferromagnets of $AE$Fe$_2$As$_2$[43] and $Ln$FeAsO.[31] Reduced moments and $G$-type ordering have also been reported for BaMn$_2$P$_2$ (4.2(1) $\mu_B$/Mn)[44] and BaMn$_2$As$_2$ (3.88(4) $\mu_B$/Mn).[43] Among the quaternary ZrCuSiAs-type compounds of Mn, the ordered moments are 3.28(5) $\mu_B$/Mn for LaMnPO (AF in the [MnP]$^-$ layer and ferromagnetic, F, between the layers),[40] 3.34(2) $\mu_B$/Mn for LaMnAsO (intralayer AF and interlayer F),[45] and ~3 $\mu_B$/Mn for PrMnSbO (AF $C$-type).[39] For these compounds, the reduced ordered moments have been attributed to a strong hybridization between pnictogen $p$ and Mn $d$ orbitals.[9,46]

Figure 8b and 8c show temperature dependence of the (1 0 1/2) magnetic peak. AF ordering temperatures of $T_N$(BaMnAsF)= 338(1) K and $T_N$(BaMnSbF)= 272(1) K are obtained from a power law fit $I(T) \propto M^2(T) \propto (1-T/T_N)^{2\beta}$ of the order parameter, with $\beta$(BaMnAsF)= 0.36(3) and $\beta$(BaMnSbF) =0.31(1), suggesting 3D Heisenberg class.[38] It is interesting that temperature dependence of the refined lattice parameters of BaMnAsF (Figure 7b) show a nonlinear trend around room temperature; this may be indicative of a magnetoelastic coupling in BaMnAsF.

**Electronic structure calculations**. All three compounds are found to be strongly magnetic with substantial Mn moments, in spite of the strong covalency between Mn and pnictogens. We perform calculations for both ferromagnetic (F) and in-plane nearest-neighbor antiferromagnetic (AF) spin configurations for the materials and consider additional magnetic configurations for BaMnAsF. In all cases, the AF configuration was lower in energy than the F configuration. The energy differences between the two configurations are 0.48 eV/f.u for BaMnAsF, 0.37 eV/f.u for BaMnSbF, and 0.33 eV/f.u for BaMnBiF. These energies are extremely high and indicate high magnetic ordering temperatures. For comparison, the prototypical ferromagnet, bcc Fe ($T_C$ = 1043 K), has a F – AF energy difference of ~ 0.4 eV in density functional calculations.[47]

The qualitative reasons for these high energies may be seen in the electronic structures. The calculated densities of states for the in-plane AF order are shown in Figure 9, with the corresponding band structures in Figure 10. As may be seen, the band structures with this magnetic order are semiconducting. The calculated gaps are 0.70 eV, 0.56 eV and 0.42 eV for BaMnAsF, BaMnSbF and BaMnBiF, respectively. The values for BaMnAsF and BaMnSbF are close to those found from transport data, which is the expected behavior of a material where the gap is between different transition metal $d$-manifolds and where Mott-Hubbard type Coulomb correlations are not large.[48] This is in contrast to simple semiconductors where DFT gaps are underestimates and especially Mott-Hubbard insulators where DFT calculations give either no gaps or gaps much smaller than experiment. The inferred relative weakness of Mott-Hubbard correlation effects in these materials is reminiscent of the FeSCs, although it should be noted that this does not mean that the FeSCs are uncorrelated.[49,50]

We find very strong covalency between the spin polarized Mn $d$ orbitals and the pnictogen $p$ orbitals in all three compounds. This is seen in the electronic densities of states for the nearest neighbor antiferromagnetic state (Figure 9). As may be seen from the projections the hybridization is strongly spin dependent, providing an explanation for the high energy scale associated with magnetic order. This is further seen by comparing the calculated electronic structures for the F and AF order. Figure 11 compares the Mn $d$ projection of the spin resolved density of states (DOS) for BaMnAsF with these two orders. There is a strong reconstruction of the electronic structure when going to the less favorable F order. In fact, this reconstruction is so strong that the semiconducting gap is closed and because of this the magnetization is reduced from the nominal value of 5 $\mu_B$/Mn to lower values of 4.0 $\mu_B$, 4.3 $\mu_B$ and 4.4 $\mu_B$ for BaMnAsF, BaMnSbF and BaMnBiF, respectively (calculated based on the total spin magnetization in the cell, not the moments in LAPW spheres). As discussed e.g. by Goodenough,[51] cases where the covalency and electronic structure are strongly affected by magnetic order are cases where high exchange couplings can be expected.

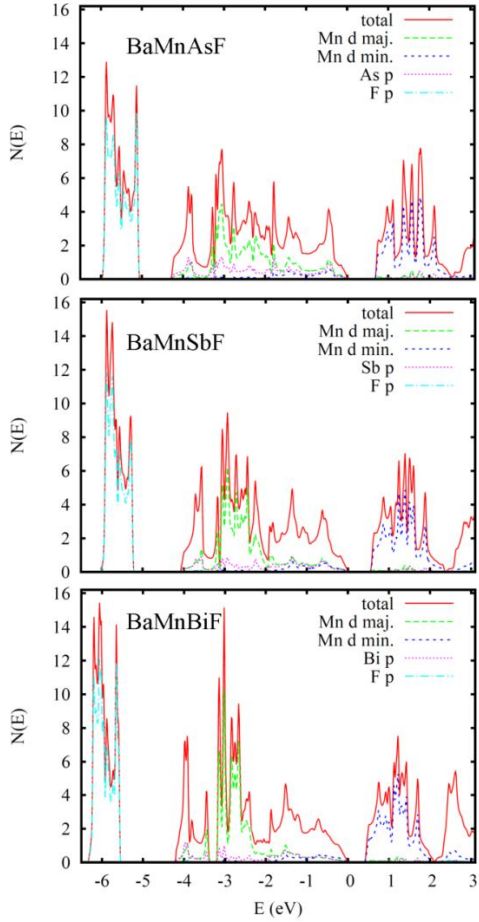

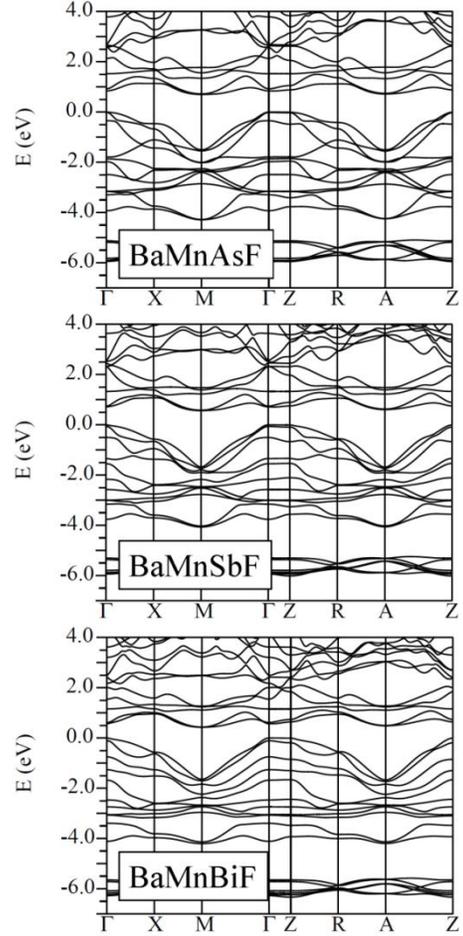

**Figure 9**. Electronic densities of states and projections for BaMn*Pn*F (*Pn* = As, Sb, Bi) in the nearest neighbor in plane antiferromagnetic state. The densities of states are per formula unit, both spins and the energy zero is at the valence band maximum.

**Figure 10**. Band structures for BaMn*Pn*F (*Pn* = As, Sb, Bi) in the nearest neighbor in plane antiferromagnetic state. The energy zero is set to the valence band maximum.

In general local moment magnetism has two ingredients: (1) moment formation and (2) the interactions between the moments that lead to order. While divalent Mn can occur in different spin states depending on the strength of the hybridization with ligands, here we find a near high spin case for all the magnetic orderings considered with similar spin moments in the Mn LAPW spheres for the F and AF cases (as well as the other cases for BaMnAsF). However, the energy associated with moment formation is not so much smaller than the ordering energy. For BaMnAsF in particular, non-spin-polarized calculations (no moments) yield an energy 1.27 eV/f.u. higher than the nearest neighbor AF order, i.e. less than three times higher than the F order (0.48 eV/f.u.).

The DOS shows that the fluorine is fully ionic occurring as $F^-$ with the *p* bands at greater than 5 eV binding energy relative to the valence band maximum, and barium is also fully ionized to $Ba^{2+}$ with Ba valence states above the conduction bands, as may be expected for such highly electropositive and electronegative atoms in proximity. This is also similar to what was found previously for BaMnPF, which was also reported to be magnetic.[14] The resulting ionic $[BaF]^+$ layers form insulating barriers between the $[MnPn]^-$ layers in the crystal structure as is seen in the relatively weak dispersion of the occupied bands along the *c*-axis (Γ-Z) in Figure 10.

Consistent with this, we find very weak magnetic interactions in the *c*-axis direction. For BaMnAsF, we calculated the energy for doubled cells along *c*-axis for the F and in-plane AF cases with and without alternation of spins along the *c*-axis. The energy differences were below 1 meV per formula unit implying that this is a very highly two dimensional magnetic system.

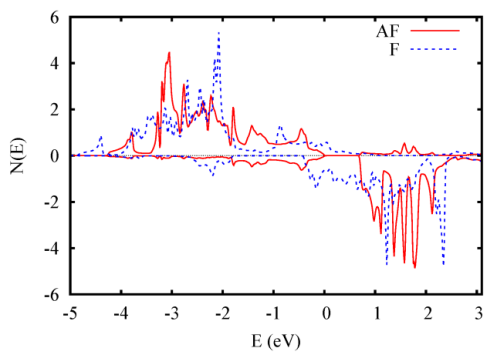

**Figure 11**. Comparison of Mn-d projected electronic DOS for BaMnAsF, with nearest neighbor in-plane antiferromagnetic ordering (AF) and ferromagnetic ordering (F). The energy zero is the valence band maximum for the AF case and the Fermi energy for the metallic F case.

In the case of BaMnAsF, we also did calculations for other AF in-plane configurations. These were for the order observed in the Fe-pnictides consisting of chains of like spin Mn atoms (the so-called SDW order) and the double stripe pattern found in FeTe[52] (see Ref. 52 for a depiction of these orders). On a per formula unit basis, we find that the SDW type ordering and the double stripe ordering are 0.12 eV and 0.21 eV higher than the nearest neighbor AF order, respectively. Therefore, in agreement with the NPD data, we conclude that the nearest neighbor in-plane order is the probable ground state.

As mentioned, we did calculations including spin orbit. In this case the energy depends on the spin orientation through the magnetocrystalline anisotropy. While this energy is small compared to the ordering energy, it is relevant to the magnetic behavior. For BaMnAsF, we did calculations with the moments oriented in the $a$-axis direction as well along the $c$-axis direction. For the other cases discussed, the calculations are done with the moments along the $c$-axis. With the experimental crystal structure the uniaxial $c$-axis direction is favored by 1.0 meV/f.u.. While this is a small energy, if correct, the result implies that there will not be a strong Kosterlitz-Thouless type reduction in the ordering temperature due to the near 2D character of the material. To summarize, the calculations find that these materials are local moment antiferromagnetic semiconductors with moderate band gaps and strong spin-dependent hybridization between the Mn $d$ states and the pnictogen $p$ states. The materials are rather two dimensional both electronically and magnetically. The strong hybridization leads to reconstructions of the band structure with changes in magnetic order. This in turn underlies very high magnetic energy scales consistent with high $T_N$ that are confirmed in neutron diffraction.

## CONCLUSIONS

Three new 1111 fluoropnictides with the ZrCuSiAs-type structure, namely BaMnAsF, BaMnSbF and BaMnBiF, are synthesized by reacting elements with $BaF_2$. The initial reactions of these components give small plate crystallites that are suitable for structure refinements using single crystal X-ray diffraction. Subsequent two-step annealing procedure results in approximately single-phase products (<2% impurity content from PXRD and magnetization results). For BaMn*Pn*F, the unit cell volumes and bond distances are much larger compared to the isostructural Mn-based oxypnictides.

Temperature dependence of electrical resistivity suggest that BaMn*Pn*F are semiconductors with band gaps of $E_g$(BaMnAsF)= 0.73 eV, $E_g$(BaMnSbF)= 0.48 eV, and $E_g$(BaMnBiF)= 0.003 eV. These values are comparable to the theoretical gaps of 0.70 eV for BaMnAsF and 0.56 eV for BaMnSbF, but not 0.42 eV for BaMnBiF. The large discrepancy in the band gap for BaMnBiF derived from the resistivity data is likely due to the influence of doping. Based on electronic structure calculations, BaMn*Pn*F are strongly magnetic with a preferred *G*-type antiferromagnetic ground state. Neutron powder diffraction (NPD) results give evidence for the *G*-type antiferromagnetic order below $T_N$ = 338(1) K for BaMnAsF, and $T_N$ = 272(1) K for BaMnSbF. However, temperature dependence of magnetization $\chi(T)$ data on these polycrystalline pellets show no anomalies at $T_N$, and $\chi(T)$ increase with increasing temperature above $T_N$. Similar featureless $\chi(T)$ data and non-Curie-Weiss behavior was recently reported for LaMnPO,[40] and increasing $\chi(T)$ above ordering temperatures are observed for $BaMn_2As_2$,[41] $BaMn_2Bi_2$,[35] $BaFe_2As_2$,[3] and LaFeAsO.[42] Such behavior may be indicative of strong exchange coupling among Mn moments in BaMn*Pn*F. Another important feature of BaMn*Pn*F phases is a strong hybridization between Mn and *Pn* states, which is responsible for reduced ordered moment values of 3.65(5) $\mu_B$/Mn for BaMnAsF, and 3.66(3) $\mu_B$/Mn for BaMnSbF, as determined from the NPD data at 4 K. The observed semiconducting antiferromagnetic behavior of BaMn*Pn*F (*Pn* = As, Sb and Bi) is similar to the reported behavior in Mn-based oxypnictides *Ln*Mn*Pn*O (*Ln* = La-Sm; *Pn* = P, As).[10]

In conclusion, further studies of transition-metal-based ZrCuSiAs-type compounds in general, and BaMn*Pn*F in particular, are warranted. Although electronic structures of BaMn*Pn*F are not similar to that of the superconducting Fe-based 1111 compounds, recent reports show that 1111 phases may demonstrate interesting variation of electrical and magnetic properties with doping and under applied pressure; for example, antiferromagnetic insulator LaMnAsO turns ferromagnetic metal when doped with hydrogen.[53] Another Mn-based compound, LaMnPO, transforms from an AF insulator to an AF correlated metal under pressure.[40] In addition, there is continued interest in 1111 phases due to the recent discovery of a high thermoe-lectric efficiency in BiCuSeO, which is further enhanced by introduction of Cu defects.[36]

## AUTHOR INFORMATION

### Corresponding Author

* E-mail: saparovbi@ornl.gov

### Notes

The authors declare no competing financial interest.

## ACKNOWLEDGMENT

This work was supported by the Department of Energy, Basic Energy Sciences, Materials Sciences and Engineering Division. Work at the High Flux Isotope Reactor, Oak Ridge National Laboratory, was sponsored by the Scientific User Facilities Division, Office of Basic Energy Sciences, US Department of Energy. We thank R. Custelcean for his help with the single crystal X-ray diffraction measurements.

Supporting Information

**Table S1** Rietveld refinement results of room temperature powder X-ray diffraction data for BaMnAsF, BaMnSbF and BaMnBiF, with *P*4/*nmm*.

| Atom | Wyckoff site | x | y | z | $U_{eq}$ (Å$^2$) |
|------|--------------|------|------|-----------|-----------|
| \multicolumn{6}{c}{BaMnAsF} | | | | | |
| \multicolumn{6}{c}{a = 4.2739(1) Å, c = 9.5875(2) Å} | | | | | |
| Ba | 2c | 0.25 | 0.25 | 0.3419(3) | 0.026(1) |
| Mn | 2a | 0.25 | 0.75 | 0 | 0.018(3) |
| As | 2c | 0.25 | 0.25 | 0.8447(5) | 0.010(2) |
| F | 2b | 0.25 | 0.75 | 0.5 | 0.014(7) |
| \multicolumn{6}{c}{BaMnSbF} | | | | | |
| \multicolumn{6}{c}{a = 4.4791(1) Å, c = 9.8297(2) Å} | | | | | |
| Ba | 2c | 0.25 | 0.25 | 0.3541(2) | 0.024(2) |
| Mn | 2a | 0.25 | 0.75 | 0 | 0.022(3) |
| Sb | 2c | 0.25 | 0.25 | 0.8321(2) | 0.021(2) |
| F | 2b | 0.25 | 0.75 | 0.5 | 0.029(8) |
| \multicolumn{6}{c}{BaMnBiF} | | | | | |
| \multicolumn{6}{c}{a = 4.5384(1) Å, c = 9.8929(2) Å} | | | | | |
| Ba | 2c | 0.25 | 0.25 | 0.3578(2) | 0.018(1) |
| Mn | 2a | 0.25 | 0.75 | 0 | 0.030(2) |
| Bi | 2c | 0.25 | 0.25 | 0.8244(2) | 0.024(1) |
| F | 2b | 0.25 | 0.75 | 0.5 | 0.017(6) |

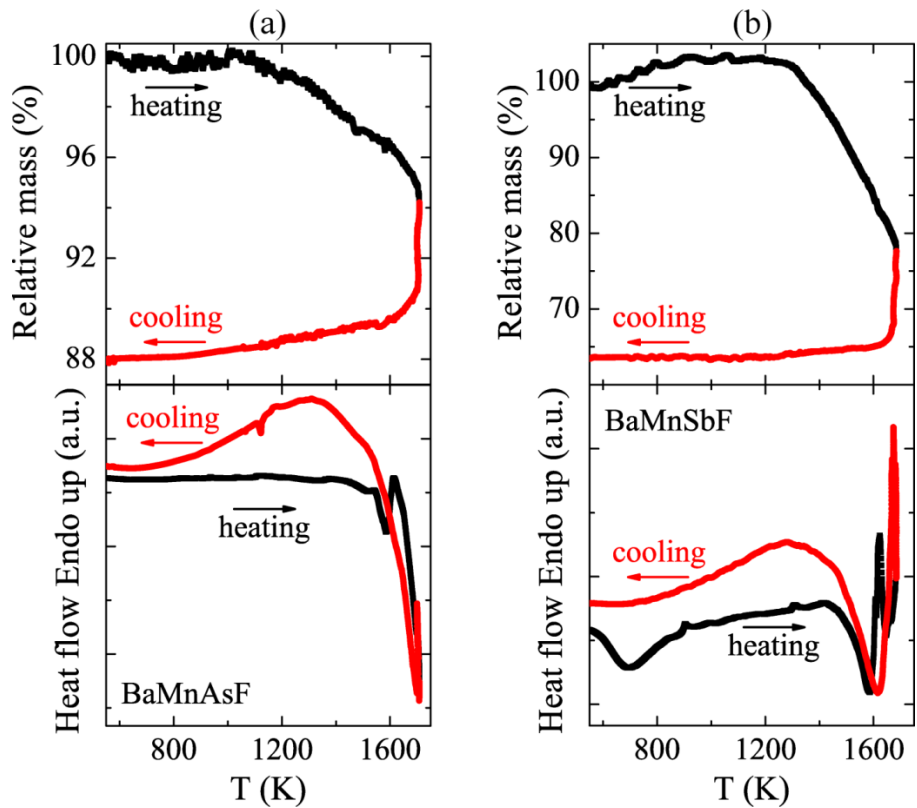

**Figure S1**. TGA and DTA data for (a) BaMnAsF and (b) BaMnSbF.

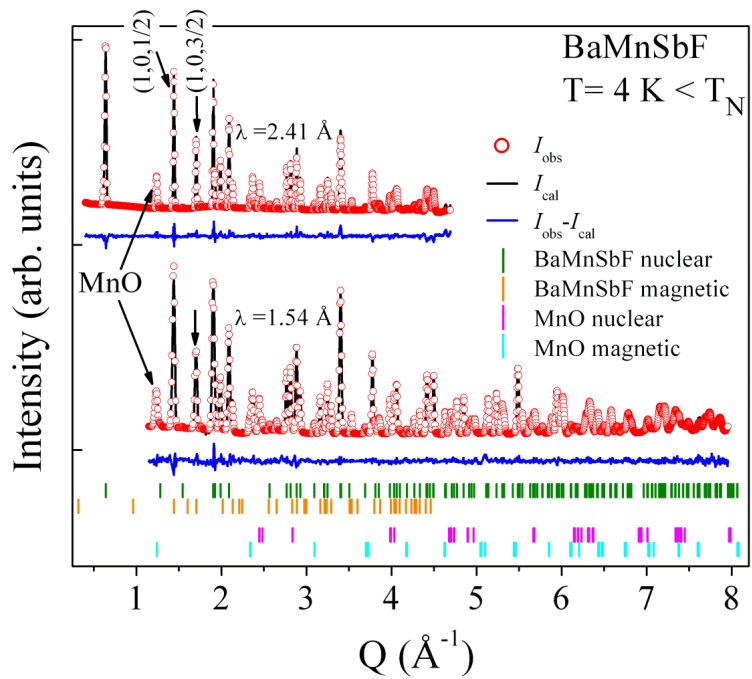

**Figure S2**. Rietveld plots of the NPD data for BaMnSbF collected at 4 K using 1.54 Å and 2.41 Å wavelengths. The two most prominent magnetic reflections, indexed as (1 0 1/2) and (1 0 3/2), are indicated by arrows. Contribution to patterns in weight fraction are BaMnSbF: 95.2(1.7)%; $BaF_2$: 2.46(0.03)%; $Mn_{1-x}O$: 2.4(0.2)%. Note: the MnO is monoclinic at low temperatures. Structural distortion is shown to appear in the non-stoichiometric MnO.

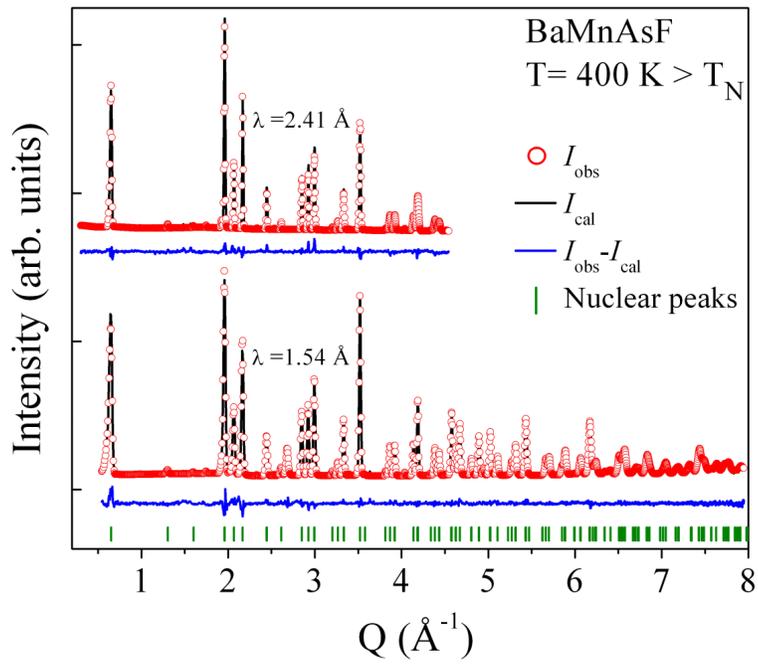

**Figure S3**. Rietveld plots of the NPD data for BaMnAsF collected at 400 K using 1.54 Å and 2.41 Å wavelengths. Only nuclear Bragg peaks are observed.

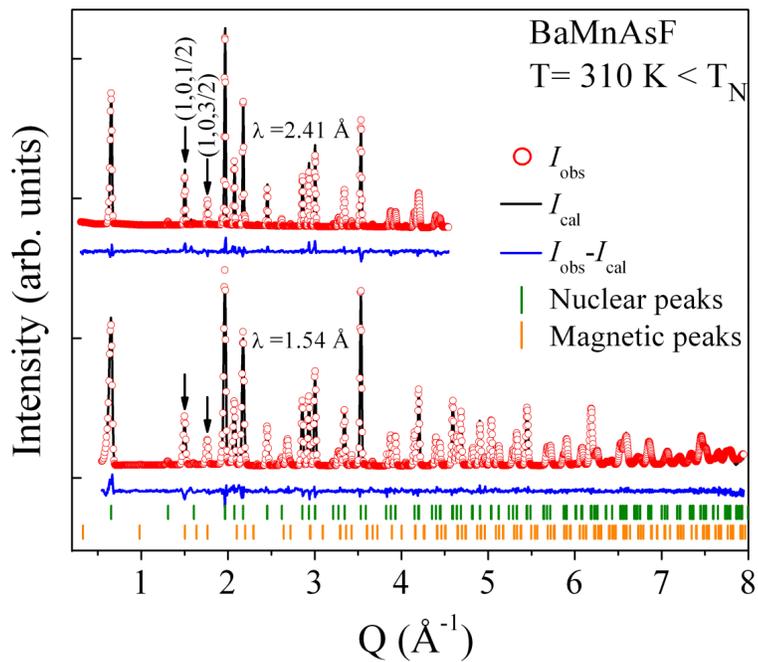

**Figure S4**. Rietveld plots of the NPD data for BaMnAsF collected at 310 K using 1.54 Å and 2.41 Å wavelengths. The two most prominent magnetic reflections, indexed as (1 0 1/2) and (1 0 3/2), are marked by arrows.